\documentclass[aps,prl,twocolumn,preprintnumbers,10pt,showpacs]{revtex4-1}
\usepackage{amsmath,bm,amssymb}

\usepackage{graphicx}
\usepackage{subcaption}
\usepackage{psfrag}
\usepackage{xcolor}
\usepackage{soul} 
 \usepackage[utf8]{inputenc}

\newcommand\eps{\epsilon}

\def\cO{  {\cal O}  }

\usepackage{slashed}

\allowdisplaybreaks

\begin{document}

\preprint{MPP-2018-306, ZU-TH 50/18, MITP/19-004}

\title{All master integrals for three-jet production at NNLO}

\author{D.\ Chicherin$^{a}$, T.\ Gehrmann$^{b}$, J.\ M.\ Henn$^{a}$, P.\ Wasser$^{c}$, Y.\ Zhang$^{a}$, S.\ Zoia$^{a}$}

\affiliation{
$^a$ Max-Planck-Institut f{\"u}r Physik, Werner-Heisenberg-Institut, 80805 M{\"u}nchen, Germany\\
$^b$ Physik-Institut, Universit{\"a}t Z{\"u}rich, Wintherturerstrasse 190, CH-8057 Z{\"u}rich, Switzerland\\
$^c$ PRISMA Cluster of Excellence, Institute of Physics, Johannes Gutenberg University, D-55099 Mainz, Germany
}

\begin{abstract}
We evaluate analytically all previously unknown nonplanar master integrals for massless five-particle scattering at two loops, 
using the differential equations method. A canonical form of the differential equations is obtained by identifying integrals with 
constant leading singularities, in $D$ space-time dimensions. These integrals evaluate to 
 $\mathbb{Q}$-linear combinations of multiple polylogarithms of uniform weight at each order in the expansion in the dimensional regularization parameter, 
 and are in agreement with previous conjectures for nonplanar pentagon functions. Our results provide the complete set of 
 two-loop Feynman integrals for any massless 
  $2\to 3$ scattering process, thereby opening up a new level of precision collider phenomenology. 
\end{abstract}

\maketitle

\section{Introduction}

The ever improving experimental precision at the LHC challenges theoretical physicists to keep up with the accuracy of the corresponding theoretical predictions. In order for this to be possible, analytic expressions for higher-loop amplitudes play a crucial role.
Among the processes that are investigated at hadron colliders, jet production observables offer unique opportunities for 
precision measurements. In particular, 
the
ratio of three- and two-jet cross sections gives a measure of the strong coupling constant $\alpha_S(Q^2)$ at high energy scales 
$Q^2$~\cite{Bendavid:2018nar,ATLAS:2015yaa,Aaboud:2017fml,Aaboud:2018hie,CMS:2014mna,Chatrchyan:2013txa}. 

While many results for next-to-next-to leading order (NNLO) cross
sections are available for $2\to 2$  processes, 
 higher multiplicity reactions
are only beginning to be explored \cite{Badger:2013gxa,
  Badger:2015lda, Gehrmann:2015bfy, Badger:2017jhb, Abreu:2017hqn, Boels:2018nrr,
  Abreu:2018jgq, Badger:2018enw, Abreu:2018zmy}, so far mostly in the planar limit. 

The situation was somewhat similar about fifteen years ago at NLO, when
novel theoretical ideas led to what is now called the ``NLO
revolution''~\cite{Ellis:2011cr}. Thanks to recent progress in quantum field theory
methods, we are today at the brink of an NNLO revolution. 

The new ideas include cutting-edge integral reduction techniques based on finite fields and algebraic geometry~\cite{vonManteuffel:2014ixa, Peraro:2016wsq, Zhang:2012ce},
a systematic mathematical understanding of special functions appearing in Feynman integrals~\cite{Goncharov:2010jf,Duhr:2011zq}, and their computation via 
differential equations~\cite{Gehrmann:1999as} in the canonical form~\cite{Henn:2013pwa}. 
The latter in fact lead to simple iterated integral solutions that have
uniform transcendental weight (UT), also called pure functions. 

It is particularly interesting that many properties of the integrated
functions can be anticipated from properties of the simpler
Feynman loop integrands through the study of the so-called leading singularities~\cite{ArkaniHamed:2010gh}.
A useful conjecture~\cite{ArkaniHamed:2010gh, Henn:2013pwa} allows one to predict which Feynman integrals satisfy the canonical differential equation by analyzing their four-dimensional leading singularities. This can be done
algorithmically \cite{WasserMSc}. 

It turns out that in complicated cases, especially
when many scales are involved, the difference between treating the
integrand as four- or $D$-dimensional can become
relevant. In particular, integrands whose numerators contain Gram determinants that vanish in four dimensions may spoil the UT property. 

In this Letter we propose a new, refined criterion for finding the
canonical form of the differential equations, and hence UT
integrals. The method involves computing leading singularities in
Baikov representation \cite{Baikov:1996rk}. 

We apply our novel technique to the most complicated nonplanar massless
five-particle integrals at NNLO. We explain how the UT basis is
obtained, and derive the canonical differential equation. We determine analytically the boundary values  
 by requiring physical consistency. 
The solutions are found to be in agreement with a previous conjecture
for nonplanar pentagon functions, and also
with a previously conjectured second entry condition~\cite{Chicherin:2017dob}.

This result completes the analytic calculation of all master integrals required for three-jet production at hadron colliders to 
NNLO in QCD. We expect that our method will have
many applications for multi-jet calculations in the near future.

\section{Integral families}

\begin{figure}
    \centering
    \begin{subfigure}[b]{0.14\textwidth}
        \includegraphics[width=\textwidth]{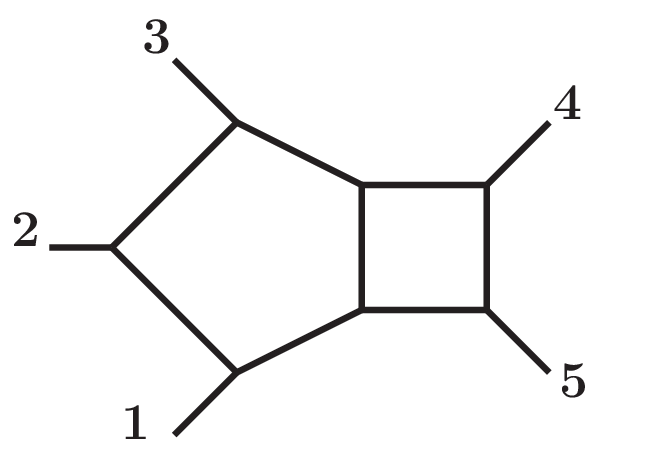}
        \caption{\\penta-box}
        \label{fig:pb}
    \end{subfigure}
    \begin{subfigure}[b]{0.15\textwidth}
        \includegraphics[width=\textwidth]{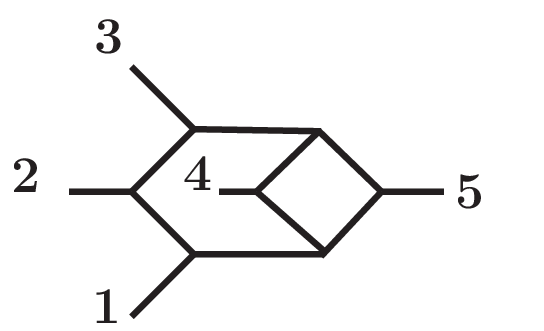}
        \caption{\\hexa-box}
        \label{fig:xt}
    \end{subfigure}
    \begin{subfigure}[b]{0.15\textwidth}
        \includegraphics[width=\textwidth]{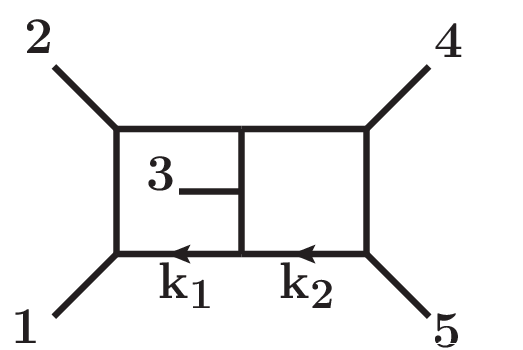}
        \caption{double-pentagon}
        \label{fig:xb}
    \end{subfigure}
    \caption{Integral topologies for massless five-particle scattering at two loops.}\label{fig:2l5pTopo}
\end{figure}

Figure~\ref{fig:2l5pTopo} shows the integral topologies needed for studying the scattering of five massless particles at two loops.
The master integrals of the planar topology shown in Fig.~\ref{fig:pb} were computed in Ref.~\cite{Gehrmann:2015bfy,Papadopoulos:2015jft,Gehrmann:2018yef}. The nonplanar integral family depicted in Fig.~\ref{fig:xt}
was computed in~\cite{Chicherin:2018mue}. (See also~\cite{Chicherin:2017dob,Chicherin:2018ubl,Chicherin:2018wes,Abreu:2018rcw}). In
this Letter, we compute the previously unknown master integrals of the double-pentagon family shown in Fig.~\ref{fig:xb}.

Genuine five-point functions depend on five independent Mandelstam
invariants, $ X = \{ s_{12} \,, s_{23} \,, s_{34} \,, s_{45} \,, s_{15} \}$, 
where $s_{ij}=2 p_{i} \cdot p_{j}$, and $p_i$ are massless external momenta.
We also introduce the parity-odd invariant $\epsilon_5$ as
\begin{align}
\label{eq:eps5}
 \epsilon_5 = \text{tr}\left[\gamma_5 \slashed{p}_1 \slashed{p}_2
  \slashed{p}_3 \slashed{p}_4 \right] \,.
\end{align}
We denote the loop momenta for the double-pentagon family by $k_1$ and $k_2$, defined as shown in Fig.~\ref{fig:xb}.

The inverse propagators are
\begin{gather}
  \begin{array}{ll}
D_1=k_1^2 \,,& D_2=(-p_1 + k_1)^2\,,\\
 D_3=(-p_1 - p_2 + k_1)^2\,, & D_4= k_2^2\,,\\
 D_5=(p_4+p_5 + k_2)^2\,, & D_6=(p_5+ k_2)^2\,, \\
D_7=(k_1 - k_2)^2\,, &  D_8=(p_3 + k_1 - k_2)^2\,, \\
D_9=(p_5+ k_1)^2 \,, & D_{10}=(-p_1 + k_2)^2\,,  \\
D_{11}=(-p_1-p_2 + k_2)^2 \,,
 \end{array}
\end{gather}
where $D_9$, $D_{10}$ and $D_{11}$ are irreducible scalar products
(ISPs).

\section{Leading singularities and uniform transcendental weight integrals}
The integrals of the double-pentagon family, shown in
Fig.~\ref{fig:xb}, can be related through integration-by-parts
relations \cite{Smirnov:2008iw, vonManteuffel:2012np, Georgoudis:2016wff} to a basis of $108$ master integrals. Out of these, $9$ are in the so-called top sector, namely they have all $8$ possible propagators. Our goal is therefore to find $108$ linearly independent UT integrals. 

The integrals of the sub-topologies are already known, because they are either sub-topologies of the penta-box~\cite{Gehrmann:2015bfy,Gehrmann:2018yef} and of the hexa-box~\cite{Chicherin:2018mue} families, or they correspond to sectors with less than five external momenta~\cite{Gehrmann:2000zt,Gehrmann:2001ck}. In order to complete the UT basis, we begin by searching for 
four-dimensional $d\log$ integrals, which are closely related to UT integrals~\cite{ArkaniHamed:2010gh}. 

An $\ell$-loop four-dimensional $d \log$ integral is an integral whose four-dimensional integrand $\Omega$ can be cast in the form 
\begin{align}
\label{eq:dlog_form}
\Omega = \sum_{I=(i_1,...,i_{4\ell})} c_I \, d \log R_{i_1} \wedge \ldots \wedge d\log R_{i_{4\ell}} \, ,
\end{align}
where the $\mathbb{Q}$-valued constants $c_I$ are the leading singularities of $\Omega$. 

In order to perform the loop integration in $D=4- 2\eps$ dimensions, where $\eps$ is the dimensional regulator,
it is necessary to clarify how the integrand is to be defined away from four dimensions. 
For example, one may simply ``upgrade" the loop momenta from 4-dimensional 
 to $D$-dimensional (abbreviated as $4d$ and $Dd$) ones.
 We call this the ``na\"ive upgrade'' of a $4d$ integrand.
 While this method is quite powerful in finding a UT basis, and indeed it has already found many successful applications~\cite{Henn:2013pwa,Henn:2014qga},
 the freedom involved in the upgrade can become important, especially for integrals with many kinematic scales.
 We first review the four-dimensional analysis, and then provide a method of fixing the freedom,
while maintaining the advantages of the canonical differential equations method. 

In this Letter, we use two techniques to find $4d$ $d\log$ integrals.

(1) The algorithm~\cite{WasserMSc}, which can decide if a given rational integrand can be cast in $d\log$ form~\eqref{eq:dlog_form}. Starting from a generic ansatz for the numerator, this algorithm can classify all possible $4d$ $d\log$ integrals in a given family.

(2) Using computational algebraic geometry, we consider a generic ansatz for the numerator $N_{\text{even}}=\sum_{\alpha} c_\alpha m_\alpha$ of the parity-even,
 or $N_{\text{odd}}=\sum_{\alpha} c_\alpha m_\alpha/\epsilon_5$ of the parity-odd $d\log$ integrals. Each $c_\alpha$ is a polynomial in $s_{ij}$, and
 $m_\alpha$ is a monomial in the scalar products. By requiring the
 $4d$ leading
 singularities of the ansatz to match a given list of rational numbers, we can use the module lift techniques~\cite{Greuel:2007:SIC:1557288} in
 computational algebraic geometry to calculate all $c_\alpha$ and to obtain a
 $4d$ $d\log$ basis. This method usually needs only a very simple
 ansatz, and the module lift can then be performed through the computer algebra system {\sc Singular}~\cite{DGPS}.

One interesting phenomenon is that, for the double-pentagon family, the na\"ive upgrade of a $4d$ $d\log$ integral is in general not UT. Let us take the $4d$ $d\log$ integrals presented in Ref.~\cite{Bern:2015ple} as examples.  
The sum of the first and the fifth $d\log$ integral numerators for the double-pentagon diagram in Ref.~\cite{Bern:2015ple}, which we denote by $B_1+B_5$, does not yield a UT integral after the na\"ive upgrade. This can be assessed from the explicit computation of the differential equation. 

The obstruction of the na\"ive upgrade implies that, in order to obtain UT integrals, we have to consider terms in the integrands which vanish
as $D=4$. These terms can be conveniently constructed from Gram determinants involving the loop momenta $k_1$ and $k_2$,
\begin{equation}
  \label{Gram}
  G_{ij}= G\left(\begin{array}{ccccc}
         k_i,p_1,p_2, p_3 , p_4 \\
         k_j,p_1,p_2, p_3 , p_4
        \end{array} \right) ,  \quad \text{with} \ i,j \in \{1,2\} \,.
\end{equation}
An integrand whose numerator is proportional to a combination of the different $G_{ij}$ explicitly vanishes in the $D\to 4$ limit. UT integral criteria based on $4d$ cuts or $4d$ $d\log$ constructions can not detect these Gram determinants, and may yield inaccurate answers on whether an integral is UT in $D$ dimensions or not. 

Instead, we develop a new $D$-dimensional criterion for UT integrals, based on the study of the cuts in Baikov representation. Our method analyzes the
$Dd$ leading singularities, and for a given $4d$ $d\log$ integral
with $4d$ integrand $N/(D_1 \ldots D_k)$, our criterion generates
a $Dd$ integrand of the form
\begin{equation}
  \label{upgrade}
  \frac{\tilde N}{\tilde D_1 \ldots \tilde D_k} + \frac{\tilde S}{\tilde D_1 \ldots \tilde D_k}\,,
\end{equation}
which is a UT integral candidate. Here the tilde sign denotes the na\"ive upgrade, and $\tilde S$ is proportional to Gram determinants. We name Eq.~\eqref{upgrade} the refined upgrade of the $4d$ $d\log$ integrand $N/(D_1 \ldots D_k)$.
The details of this $D$-dimensional criterion based on Baikov cuts are given in the next section. 

Applying our method to the top sector of the double-pentagon family leads to two observations.

(1) For any $4d$ double-pentagon $d\log$ in Ref.~\cite{Bern:2015ple} we can find its refined upgrade from our $Dd$ UT criterion. We verified that such refined upgrades are indeed UT integrals by computing the differential equation. For example, the refined upgrade of $(B_1+B_5)$ is
\begin{align}
 &(\tilde{B}_1+\tilde{B}_5)+\frac{16 s_{45}G_{12} }{\epsilon_5^2} \times \nonumber \\
&(s_{12} s_{23}-s_{12} s_{15}  + 2 s_{12} s_{34} + s_{23} s_{34} + s_{15} s_{45} - s_{34} s_{45})\,.
\label{xb101}
 \end{align}
 
 (2) Some integrals with purely Gram determinant numerators satisfy our $Dd$ UT criterion:
\begin{align}
&\frac{s_{45}}{\epsilon_5} (G_{11}-G_{12}),\ \ 
\frac{s_{12}}{\epsilon_5} (G_{22}-G_{12}), \ \
\frac{s_{12}-s_{45}}{\epsilon_5} G_{12}\label{gram_UT} \,.
\end{align}
Once again we verified that these integrals are indeed UT by examining the differential equation.

\section{Criterion for pure integrals from $D$-dimensional cuts}
In this section we present our new criterion for UT integrals based on $Dd$ cuts in the Baikov representation~\cite{Baikov:1996rk}. As we have already seen, this new criterion is sharper than the 
original $4d$ one, as it can also detect Gram determinants to which the latter is blind.

Let us recall that in the Baikov representation \cite{Baikov:1996rk} the propagators of a $Dd$ Feynman integrand are taken to be integration variables
(Baikov variables). The $Dd$ leading singularities can thus
be calculated easily by taking iterative residues. 
Then, our $Dd$ criterion for a UT integral is to require all the residues of its Baikov representation to be rational numbers. 

For the double-pentagon integral family, the standard Baikov cut analysis~\cite{Bosma:2017ens,
  Harley:2017qut}, based on the two-loop Baikov representation, eventually leads to complicated three-fold integrals. To avoid this computational
difficulty, we adopt the loop-by-loop Baikov cut analysis~\cite{Frellesvig:2017aai}.

For a double-pentagon integral with some numerator $N$, for instance, the integration can be separated loop-by-loop as
\begin{align}
\label{baikov-loop-by-loop-G12}
I_{\text{dp}}[N]
= \int d^D k_2
  \frac{1}{D_4 D_5 D_6}\int d^D k_1 \frac{N}{D_1 D_2 D_3 D_7 D_8} \,.
\end{align} 
The two-loop integral can thus be decomposed into a pentagon integral with loop momentum $k_1$ and external legs $p_1$, $p_2$, $p_3$ and $-k_2$, and a
triangle integral with loop momentum $k_2$. Note that, if necessary, we might need to carry out a one-loop integrand reduction for the numerator $N$ first, in order to make sure that the integrand contains no cross terms such as $k_1 \cdot p_4$ or $k_1 \cdot p_5$. As a consequence, $D_9$ drops out from the integrand.

We then apply the Baikov representation loop-by-loop, i.e.\ we change integration variables from the components of the loop momenta to $10$ Baikov variables, $z_i\equiv D_i$, $i\in \{1,\ldots, 11\}\backslash
\{9\}$. Once this is done, we can explore the $Dd$ residues.

For instance, consider the double-pentagon integral $I_\text{dp}[G_{12}]$. Its $4d$ leading singularities are all vanishing, and can therefore not
determine whether $I_\text{dp}[G_{12}]$ is UT or not. Conversely, by using our Baikov cut method, having integrated out the term $k_1 \cdot
p_4$, we get a Baikov integration with $10$ variables. 
Taking the residues in $z_i = 0$, $\forall i\in C$, where $C \subseteq
 \{1,\ldots,8\} $, yields integrands which do not vanish in the $D\to 4$ limit. Using
 the algorithm~\cite{WasserMSc}, we systematically compute all possible
 residues of these integrands in the remaining variables, and make sure
 that there are no double poles. In this way we compute the leading
 singularities on different cuts, and find that they all evaluate to $\pm
 \epsilon_5/(s_{12}-s_{45})$ or zero. As a result, we see that the integral
 \begin{equation} \label{G_12_UT} \frac{s_{12}-s_{45}}{\epsilon_5}
   I_\text{dp}[G_{12}] 
\end{equation} 
satisfies our $Dd$ criterion. We confirmed that \eqref{G_12_UT} is indeed a UT integral by explicitly computing it from differential equations.

Similarly, we can use this loop-by-loop Baikov cut method to
find the UT integral candidates listed in Eqs.~\eqref{xb101} and~\eqref{gram_UT}, for which the $4d$ leading singularity calculation cannot give a definitive answer. All these candidates 
are subsequently proven to be UT by the differential equations. 

It is worth noting that this $Dd$ Baikov cut analysis only
involves basic integrand reduction and residue computations. We expect that
this method, combined with the $d\log$ construction algorithm described in~\cite{WasserMSc}, will prove to be a highly efficient way of determining UT integral candidates for even more complicated diagrams
in the future.

\section{Master integrals and  canonical differential equations}
With the study of $4d$ $d\log$ integrals, and the novel $Dd$ Baikov cut
analysis, we constructed a candidate UT integral basis for
the double-pentagon family. 

Through IBPs, we find that the eight $4d$ $d\log$s in Ref.~\cite{Bern:2015ple}, after our refined upgrade, together with the three Gram-determinant integrals
given in Eq.~\eqref{gram_UT}, span a $8$-dimensional linear space. By combining the algorithm described in~\cite{WasserMSc} and the computational algebraic geometry method, we easily find another linearly independent integral satisfying our $Dd$ UT criterion. This completes the basis for the double-pentagon on the top sector. Sub-sector UT integrals are either found via~\cite{WasserMSc}, or taken from the literature~\cite{Gehrmann:2015bfy,Gehrmann:2018yef,Chicherin:2018mue}.

By differentiating our candidate UT basis for the double-pentagon family, we see that the differential equations are immediately in the canonical form~\cite{Henn:2013pwa}
\begin{align}
\label{canonicalDEpentagon}
d \vec{I}(s_{ij};\epsilon) \,=\, \epsilon \, d \tilde{A} (s_{ij}) \, \vec{I}(s_{ij};\epsilon) \,,
\end{align} 
without the need for any further basis change. This is the ultimate proof that our basis integrals are indeed UT.

We wish to emphasize here that the construction of the UT basis
is done at the integrand level via Baikov cut analysis, and as
such does not require the a priori knowledge of the differential
equations. 

It is also worth mentioning that the analytic inverse of the transformation matrix between our UT basis and the ``traditional" basis from Laporta algorithm was efficiently computed by means of the sparse linear algebra techniques described in~\cite{Boehm:2018fpv}.

Equation~\eqref{canonicalDEpentagon} can be further structured to the form
\begin{align}
\label{canonicalDEpentagon_letter}
d\vec{I}(s_{ij};\epsilon) = \epsilon \left( \sum_{k=1}^{31} a_k d\log W_k(s_{ij}) \right) \vec{I}(s_{ij};\epsilon) \, ,
\end{align}
where $W_k$ are letters of the pentagon symbol alphabet conjectured in~\cite{Chicherin:2017dob}, and each $a_k$ is a $108\times
108$ rational number matrix.

We consider the integrals in the $s_{12}$ scattering region. The latter is defined by positive $s$-channel energies,
$ \{s_{12}, s_{34}, s_{45}, s_{35} \}\ge0$, and negative $t$-channel energies, $\{s_{23}, s_{24},s_{25}, s_{13},s_{14},s_{15} \} \le 0$, as well as reality of particle momenta, which translates to $\Delta \le 0$.

We choose a boundary point 
\begin{align}
X_{0} = \{ 3, -1, 1,1,-1 \}  
\end{align}
 inside this region. 
We determine the boundary values of the integrals
by requiring physical consistency, as described in~\cite{Chicherin:2018mue}. 
This yields a system of equations for the boundary constants at $X_{0}$, whose
coefficients are Goncharov polylogarithms. We evaluate the latter to high precision
using GiNaC~\cite{Bauer:2000cp}.
The values at $X_{0}$ were validated successfully with the help of SecDec~\cite{Borowka:2015mxa}.

The full result for the integrals is again written in terms of Goncharov polylogarithms.
For reference, we provide numerical values for all integrals at the symmetric point $X_0$, as well as
for an asymetric point 
\begin{align}
X_1 = \left\{ 4,-\frac{113}{47},\frac{281}{149},\frac{349}{257},-\frac{863}{541}  \right\}\,.
\end{align}
The values,  given in ancillary files, have at least $50$ digit precision.
Here we display the results for integral $I_{107}$,
\begin{align}
& I_{107}(X_0 , \eps) = 16.383606637078885171 i+ \cO(\eps) \,,\\
& I_{107}(X_1 , \eps) = 6.9362922441923047974 i + \cO(\eps) \,.
\end{align}
From their leading order term in $\epsilon$ of the boundary values, one can immediately write down the symbol of the integrals. This has also 
been computed independently in~\cite{Abreu:2018aqd}, and 
has already been employed in the computation of two-loop five-point amplitudes in $\mathcal{N}=4$ super-Yang-Mills theory~\cite{Abreu:2018aqd,Chicherin:2018yne} and $\mathcal{N}=8$ supergravity~\cite{Chicherin:2019xeg,Abreu:2019rpt} at symbol level.
We observe that the second entry condition conjectured in~\cite{Chicherin:2017dob} is indeed satisfied.

We provide the UT basis for the double-pentagon family, the $\widetilde{\mathcal{A}}$ matrix of the canonical differential equation~\eqref{canonicalDEpentagon}, and the boundary values at $X_{0}$ and $X_{1}$ in ancillary files.

\section{Discussion and Outlook}
In this Letter, we computed analytically the master integrals of the last missing integral family needed for massless five-particle scattering amplitudes at two loops. We applied the canonical differential equation method~\cite{Henn:2013pwa}, supplemented with a novel strategy for finding integrals evaluating to pure functions based on the analysis of $Dd$ leading singularities in Baikov representation.

Our calculation confirms the previously conjectured pentagon functions alphabet and second entry condition~\cite{Chicherin:2017dob}. Our result implies the latter is a property of individual Feynman integrals, not only of full amplitudes. It will be interesting to find a field theory explanation of this condition, perhaps along the lines of the Steinmann relations.

With our result, all master integrals relevant for three-jet production at NNLO are now known analytically. 
Moreover, they are ready for numerical evaluation in physical scattering regions.
This opens the door to computing full $2\rightarrow 3$ scattering amplitudes at two loops.

We expect that our $Dd$ Baikov cut analysis will prove to be a powerful method to find Feynman integrals evaluating to pure functions, in particular for integral families involving many scales. 
We expect it will have many further applications for multi-particle amplitudes, e.g. for $H+2\, j$ and $V+2\,j$ productions, and other multi-scale processes relevant for collider physics.

\section*{Acknowledgments}
We are indebted to Gudrun Heinrich and Stephan Jahn for providing numerical checks and for help with using SecDec.
Y.Z. thanks Alessandro Georgoudis and Yingxuan Xu for enlightening
discussions. This research received funding from the Swiss National Science Foundation (Ambizione Grant No. PZ00P2
161341), the European Research Council (ERC) under the European Union's
Horizon 2020 Research, and Innovation Programme (Grant Agreement No. 725110), ``Novel
Structures in Scattering Amplitudes." J. H., Y. Z., and S. Z. also wish to thank the Galileo Galilei Institute for hospitality during the workshop ``Amplitudes in the LHC Era."

\bibliographystyle{h-physrev} 

\bibliography{5point_refs2.bib}
\end{document}